\documentclass[aps,prb,reprint,superscriptaddress]{revtex4-2}
\usepackage{physics}
\usepackage{enumerate}
\usepackage{graphicx}
\usepackage{amsmath}
\usepackage{amsthm}   
\usepackage{mathtools}
\usepackage{xcolor}
\usepackage{amsfonts}
\usepackage{amssymb}
\usepackage{units}
\usepackage[normalem]{ulem}

\begin{document}

\title{From Quantum to Classical Dissipative Spin Dynamics}

\author{Alireza Ataei}
\affiliation{Department of Mathematics, Uppsala University, Sweden}
\author{Leila Ataei}
\affiliation{Department of Mathematics, University of Reims Champagne-Ardenne, France}
\author{Olle Eriksson}
\affiliation{Department of Physics and Astronomy, Uppsala University, Box 516, Sweden}
\affiliation{Wallenberg Initiative Materials Science,
WISE, Uppsala University, Box 516, 
SE-751 20 Uppsala, Sweden}
\author{Vahid Azimi Mousolou}
\affiliation{Department of Physics and Astronomy, Uppsala University, Box 516, Sweden}

\date{\today}
\begin{abstract}

The Landau-Lifshitz-Gilbert (LLG) equation provides the fundamental description of dissipative magnetization dynamics. A recently proposed quantum analog, the quantum Landau-Lifshitz-Gilbert (QLLG) equation, extends this framework to quantum spin systems. However, whether the QLLG equation recovers the LLG equation in the classical large-spin limit for interacting spin systems remains an open question. Here, we demonstrate this connection by deriving the LLG equation as the classical large-spin limit of the QLLG equation. We establish this correspondence analytically and verify it numerically for an interacting spin system. These results provide a quantum-to-classical description of dissipative spin dynamics, offering a microscopic foundation for the LLG equation. From numerical calculations provided here, we also conclude that quantum effects become much less noticeable for systems with larger number of spins in the Hamiltonian.

\end{abstract}

\maketitle

\section{Introduction}

The Landau--Lifshitz--Gilbert (LLG) equation \cite{LandauLifshitz1935,Gilbert2004} provides the standard theoretical framework for describing dissipative magnetization dynamics in magnetic materials. The LLG equation has become a cornerstone of modern magnetism and provides the basis for micromagnetic modeling and atomistic spin dynamics across a broad range of magnetic systems and length scales \cite{Brown1963,Aharoni1996,Evans2014,ErikssonBook2017}. Its applicability extends from equilibrium properties and magnetic reversal to ultrafast magnetization dynamics, spin-wave propagation, and nanoscale magnetic phenomena \cite{Stanciu2007,Kirilyuk2010,Manchon2019}.

Despite its classical formulation, the microscopic degrees of freedom underlying magnetism are fundamentally quantum mechanical. In particular, localized magnetic moments originate from quantum spins, where exchange interactions, spin--orbit coupling, and magnetic anisotropies arise from the underlying electronic structure. The classical LLG description therefore represents an effective macroscopic or semiclassical description of fundamentally quantum degrees of freedom. Understanding the regime in which this description emerges from quantum spin dynamics is consequently an important aspect in magnetism \cite{Auerbach1994,KlauderSkagerstam1985}.

The role of dissipation makes this question particularly interesting. The dynamics of an isolated quantum spin system is governed by the Schr\"odinger equation, or equivalently by the Liouville--von Neumann equation for the density operator. For an open quantum system interacting with an environment, the Lindblad master equation provides the standard linear framework for describing dissipative quantum dynamics \cite{BreuerPetruccione,Gorini1976,Lindblad1976}. However, the phenomenological damping appearing in the classical LLG equation is not simply equivalent to conventional linear quantum dissipation. A microscopic description of magnetic damping can involve, for example, spin--orbit coupling, interactions with itinerant electrons, and angular-momentum transfer between spin and orbital degrees of freedom \cite{Kambersky1970,Brataas2012,Manchon2019}. These considerations motivate the development of quantum dynamical equations that retain the structure of spin dynamics while incorporating dissipation directly at the level of quantum states.

Quantum formulations of dissipative spin dynamics have in fact been proposed. In particular, quantum analogs of the Landau--Lifshitz (LL) and LLG equations have been developed using different approaches to quantum spin dynamics and damping \cite{Wieser2013,Wieser2015}. More recently, a quantum LLG (QLLG) equation has been formulated as a nonlinear, purity-preserving evolution equation for the density operator, providing a direct quantum generalization of Gilbert damping \cite{Liu2024}. Unlike conventional linear open-system dynamics, this formulation preserves the spectrum of the density operator and therefore describes an isospectral dissipative evolution. At the same time, it reduces to the classical LLG equation for a single spin-$1/2$ while retaining genuinely quantum effects in interacting many-spin systems, including the generation of entanglement and nonlocal spin correlations \cite{Liu2024,azimi2025numerical,mirzaei2025lrei}. The QLLG framework has also been proposed as a mechanism for nonlinear quantum relaxation and the preparation of low-energy and ground-state eigenspaces \cite{Ataei2026,EnergyOrdering2026}.

These developments raise a fundamental question on how the classical LLG equation emerges from the QLLG dynamics for an interacting many-spin system? Establishing such a limit is important not only for validating the QLLG framework, but also for identifying precisely how classical dissipative magnetization dynamics is embedded in the underlying quantum theory. In particular, the presence of interactions makes the question nontrivial. Even when the initial state is a product of spin-coherent states, quantum evolution can generate correlations and entanglement between different spins, potentially taking the state away from the product coherent-state manifold \cite{Liu2024,azimi2025numerical,mirzaei2025lrei,Ataei2026,EnergyOrdering2026}. A classical limit must therefore account simultaneously for the large-spin scaling of the Hamiltonian, the evolution of local spin expectation values, and the behavior of quantum fluctuations and correlations.

For non-dissipative spin dynamics, the emergence of classical equations from quantum spin systems is established \cite{ZhangBatista2021}. Spin-coherent states provide a natural semiclassical description of large spin-$S$, and coherent-state path-integral and variational methods lead to classical spin equations in the appropriate limit \cite{KlauderSkagerstam1985,Auerbach1994}. Moreover, such semiclassical descriptions and techniques provide systematic connections between quantum and classical spin dynamics \cite{Fradkin2013}.

The dissipative case is considerably nontrivial. The classical LLG equation contains a damping term that is not generated by ordinary unitary Schr\"odinger evolution, while the QLLG equation introduces dissipation through a non-unitary nonlinear quantum evolution. Consequently, the standard coherent-state derivation of classical, non-dissipative dynamics \cite{ZhangBatista2021} does not, by itself, demonstrate how Gilbert damping emerges in the classical limit of QLLG dynamics. In particular, for interacting many-spin systems one must show that the quantum correlations remain negligible throughout QLLG dynamics in the large-spin limit, allowing the local magnetizations converge to the classical LLG dynamics. To our knowledge, the classical limit of the QLLG equation for interacting many-spin systems has not yet been established.

In this work, we establish this connection by deriving the classical LLG equation as the classical large-spin limit of the QLLG equation for interacting spin systems initially prepared in product spin-coherent states. We show analytically that local quantum fluctuations remain of order $S$, such that the local magnetizations retain their classical magnitude in the large spin-S limit. We then demonstrate that the resulting quantum equations of motion converge to the classical LLG dynamics as $S\to\infty$. We further verify the correspondence numerically for model interacting spin systems by examining both the local magnetization dynamics and the suppression of quantum correlations with increasing spin number. Our results provide a quantum-to-classical correspondence for dissipative spin dynamics, establishing the QLLG equation as a consistent quantum extension of classical LLG dynamics and providing a microscopic foundation for the emergence of the classical dissipative equation.

\section{Product Coherent Spin State at Classical Limit}
\label{Product Coherent Spin State at Classical Limit}
In the quantum description, the corresponding normalized local magnetization vector is naturally defined as
\begin{equation}
\label{eq:normalizedspin}
\mathbf m_i=\Tr({\mathbf s_i} \rho),\qquad
\mathbf s_i=\frac{\mathbf S_i}{\hbar S},
\end{equation}
where $\mathbf S_i=(S_i^x,S_i^y,S_i^z)$ is the spin vector operator, and $\rho$ is the density matrix describing the state of the quantum system. The corresponding physical magnetization is then
\begin{equation}
\label{eq:magnetization}
\mathbf M_i=\Tr({\mathbf S_i} \rho)=\hbar S\,\mathbf m_i.
\end{equation}
For simplicity, and without loss of generality, we set the gyromagnetic ratio $\gamma=1$. Following the classical LLG equation \cite{LandauLifshitz1935,Gilbert2004}, the length of local magnetization vector is fixed throughout the time-evolution, and here for simplicity normalized to one;
\begin{equation}
\label{eq:classicalassum}
\|\mathbf m_i\|=1.
\end{equation}
The classical property in Eq.~\eqref{eq:classicalassum} uniquely determines the form of the quantum state. In particular, it implies that the many-body quantum state $\rho$ is a product of local spin coherent states as shown in the following. By using Eqs.~\eqref{eq:magnetization} and \eqref{eq:classicalassum}, we obtain
\begin{equation}
\hbar S
=\mathbf M_i\cdot\mathbf m_i
=\Tr\left(\rho\,\mathbf S_i\cdot\mathbf m_i\right).
\label{tracenormmagnetization}
\end{equation}
Let $R_i \in \mathrm{SO}(3)$ be a rotation, mapping the \(z\)-axis to the unit vector \(\mathbf m_i\), and
$U(R_i)$ be its unitary spin representation such that
\begin{equation}
\mathbf S_i\cdot\mathbf m_i
=
U^\dagger(R_i)\,S_i^z\,U(R_i).
\end{equation}
Substituting this in Eq. \eqref{tracenormmagnetization} yield 
\begin{equation}
\hbar S
=
\Tr\!\left(U(R_i)\rho U^\dagger(R_i)\,S_i^z\right).
\end{equation}
This implies that the expectation value of $S_i^z$ attains the upper bound, namely the largest eigenvalue $\hbar S$ of $S_i^z$. Therefore, the support of $U(R_i)\rho U^\dagger(R_i)$ lies in the eigenspace of $S_i^z$ with eigenvalue $\hbar S$. Since this eigenspace is one-dimensional, spanned by $\ket{S,S}$, the reduced state at each site $i$ is pure and given by the spin coherent state
\begin{equation}
\label{eq:respsenmagnet}
\ket{\mathbf{m}_i}=U(R_i)\ket{S,S}.
\end{equation}
This proves that the classical constraint in Eq.~\eqref{eq:classicalassum} is equivalent to the many-body state being a product of local spin coherent states, i.e.,
\begin{equation}
\label{eq:productcoherent}
\rho
=
\ket{\psi}\bra{\psi},
\qquad
\ket{\psi}
=
\ket{\mathbf{m}_1}\otimes\cdots\otimes\ket{\mathbf{m}_N}.
\end{equation}

\section{Dissipative magnetization dynamics}

\subsection{Spin dynamics}
\label{Sec:dissipativemagnetizationdynamics}
Dissipative magnetization dynamics in magnetic materials, is conventionally studied by the classical LLG equation expressed as \cite{Gilbert2004,ErikssonBook2017}
\begin{equation}
\dot{\mathbf{M}}_i
=
 \mathbf{M}_i \times \mathbf{B}_i^{\mathrm{eff}}
-  \frac{\alpha}{|\mathbf{M}_i|} 
\mathbf{M}_i \times \dot{\mathbf{M}}_i,
\label{eq:LLGeq}
\end{equation}
where $\mathbf{M}_i$ is the magnetization, $\mathbf{B}_i^{\mathrm{eff}}=-\frac{\partial H}{\partial \mathbf{M}_i}$ is the effective magnetic field given by the Hamiltonain $H$ and $\alpha$ is the damping rate. Note that the length $\|\mathbf{M}_i\|$ is preserved throughout the LLG dynamics, which is consistent with the classical fixed-length property in Eq.~\eqref{eq:classicalassum}. Thus, each local magnetic moment, $\mathbf{M}_i$, is treated as a classical vector of conserved magnitude throughout the dynamics.

Recently, a quantum analog of Eq.~\eqref{eq:LLGeq} has been proposed in Ref.~\cite{Liu2024} for a given quantum spin Hamiltonian $H$ as
\begin{equation}
\dot{\rho} = \frac{i}{\hbar}[\rho,H] + i\kappa[\rho,\dot{\rho}],
\label{QLLG}
\end{equation}
which describes the evolution of the density matrix $\rho$ in the presence of damping, with damping rate $\kappa>0$.

In the following, we derive the classical LLG equation in Eq.~\eqref{eq:LLGeq} from the QLLG equation in Eq.~\eqref{QLLG} in the classical limit 
\cite{Auerbach1994,KlauderSkagerstam1985,ZhangBatista2021},
\begin{equation}
S\rightarrow\infty,\qquad
\hbar\rightarrow0,
\label{eq:CL}
\end{equation}
with $\hbar S$ held fixed, using the initial condition specified in Eq.~\eqref{eq:classicalassum} or, equivalently, in Eq.~\eqref{eq:productcoherent}. We prove this in two steps:
\begin{enumerate} 
\item \textbf{Convergence of QLLG to classical LLG.} We show that the QLLG dynamics reduces to the classical LLG dynamics on the manifold of pure product spin-coherent states.

\item  \textbf{Coherent-state manifold dynamics.}
At the classical limit specified in Eq.~\eqref{eq:CL}, for an initial product of spin-coherent states as in Eq.~\eqref{eq:productcoherent}, the QLLG dynamics remains asymptotically within the coherent-state manifold. Explicitly, we show that 
\begin{equation} 
\label{eq:normmiunit}\|\mathbf m_i\| = \frac{\|\mathbf M_i\|}{\hbar S} = 1+O(S^{-1}) \end{equation} 
along the QLLG dynamics, where $\mathbf m_i$ and $\mathbf M_i$ are defined in Eqs.~\eqref{eq:normalizedspin} and~\eqref{eq:magnetization}, respectively.
Thus, at each time, the instantaneous local normalized magnetization approaches unit length as $S\to\infty$. By the equivalence between Eqs.~\eqref{eq:classicalassum} and \eqref{eq:productcoherent}, this establishes that the QLLG evolution becomes asymptotically confined to the manifold of product spin-coherent states. 

\end{enumerate} 

\subsection{Pure-state quantum spin dynamics}
Since the classical condition in Eq.~\eqref{eq:classicalassum} corresponds to the quantum pure state specified in Eq.~\eqref{eq:productcoherent}, only the pure-state formulation of the QLLG equation is required in the present analysis. In this case, Eq.~\eqref{QLLG} is equivalent to the nonlinear Schr\"odinger equation \cite{Ataei2026} 
\begin{equation}
\label{eq:shrodingerequation}
\partial_t \ket{\psi}
=
-\frac{i(1-i\kappa)}{\hbar(1+\kappa^2)}
\left(H - \langle H \rangle \right)\ket{\psi},
\end{equation}
where, for any observable $O$ including the Hamiltonian $H$, we define
\begin{equation}
\langle O\rangle = \bra{\psi}O\ket{\psi}.
\end{equation}
The time evolution of an observable $O$ follows directly from Eq.~\eqref{eq:shrodingerequation}, yielding
\begin{equation}
\label{eq:observableevolution}
\begin{aligned}
\frac{d}{dt}\langle O\rangle
&= 
\frac{i}{\hbar(1+\kappa^2)}\langle[H,O]\rangle
-\frac{\kappa}{\hbar(1+\kappa^2)}\mathrm{Cov}(H,O)\\&+  \big \langle \frac{d O}{dt} \big \rangle,
\end{aligned}
\end{equation}
where the symmetric covariance is defined as 
\begin{equation}
\mathrm{Cov}(O,H)
=
\langle
 \{O,H\}\rangle
-2\langle O\rangle\langle H\rangle,
\end{equation}
with $\{O,H\}$ denoting the anticommutator. Applying Eq.~\eqref{eq:observableevolution} to the local spin observable $\mathbf{S}_i$ yields
\begin{equation}
\dot{\mathbf{M}}_i
=
\frac{i}{\hbar(1+\kappa^2)}
\langle [H,\mathbf{S}_i]\rangle
-
\frac{\kappa}{\hbar(1+\kappa^2)}
\mathrm{Cov}(H,\mathbf{S}_i),
\label{eq:magnetizationevolution}
\end{equation}
for any given spin Hamiltonian $H$.

In this work, we consider a spin system whose dynamics is described by the generic spin Hamiltonian
\begin{equation}
\begin{aligned}
H &= H_B + H_{\mathcal{J}}\\&=
- \hbar \sum_i \mathbf{B}_i \cdot \mathbf{s}_i
-\frac{1}{2} \hbar^2\sum_{ij}\sum_{\alpha\beta}
s_i^\alpha \mathcal{J}_{ij}^{\alpha\beta} s_j^\beta,
\end{aligned}
\label{eq:Bilinearhamiltonian}
\end{equation}
where $\alpha,\beta= x,y,z$. For instance, the interaction tensor can be decomposed as
\begin{equation}
\mathcal{J}_{ij}^{\alpha\beta}
=
J_{ij}\delta^{\alpha\beta}
+
\varepsilon^{\alpha\beta\gamma}D_{ij}^{\gamma}
+
K_{ij}^{\alpha\beta},
\end{equation}
with $J_{ij}$, $\mathbf{D}_{ij}$, and $K_{ij}^{\alpha\beta}$ being the isotropic exchange, Dzyaloshinskii--Moriya interaction, and symmetric anisotropic exchange, respectively.

\section{Convergence of QLLG to classical LLG.}
Here, we prove Step 1 outlined in Section~\ref{Sec:dissipativemagnetizationdynamics}.
We assume that the QLLG dynamics is restricted to the manifold of product spin-coherent states, namely states of the form given in Eq.~\eqref{eq:productcoherent}. Under this assumption, we evaluate the magnetization dynamics in Eq.~\eqref{eq:magnetizationevolution} and derive its classical limit.

\subsection{Precessional evolution}

For the unitary (precessional) contribution given by the first term in Eq.~\eqref{eq:magnetizationevolution}, the product form of the state in Eq.~\eqref{eq:productcoherent} yields
\begin{equation}
\label{eq:commutator_term}
\frac{i}{\hbar}\langle [H,\mathbf{S}_i]\rangle
=
\mathbf{M}_i \times \mathbf{B}_i^{\mathrm{eff}},
\end{equation}
where the effective magnetic field corresponding to the Hamiltonian in Eq.~\eqref{eq:Bilinearhamiltonian} is given by
\begin{equation}
\label{eq:effective_field}
\mathbf{B}_i^{\mathrm{eff}}
=
\frac{1}{ S}\mathbf{B}_i
+
\frac{1}{ S^2}\sum_{j\alpha\beta}
\mathcal{J}_{ij}^{\alpha\beta} m_j^\beta \hat{e}_\alpha=-\frac{\partial \langle H\rangle}{\partial \mathbf{M}_i}.
\end{equation}
This result follows from the spin commutation relations
\begin{equation}
\label{eq:commutatorspins}
[S_i^\alpha,S_j^\beta]
=
i\hbar\delta_{ij}\varepsilon_{\alpha\beta\gamma} S_i^\gamma,
\end{equation}
where \(\varepsilon_{\alpha\beta\gamma}\) is the Levi-Civita symbol and the fact that 
\begin{equation}
\label{eq:commutatorexpectation}
\langle S_k^\beta S_i^\alpha \rangle
=
\langle S_k^\beta\rangle \langle S_i^\alpha\rangle,
\qquad k \neq i,
\end{equation}
for a local product state. Note that Eq.~\eqref{eq:commutator_term} does not require the local states to be coherent.

\subsection{Dissipative evolution}

The second dissipative covariance term in Eq.~\eqref{eq:magnetizationevolution} can be decomposed into \begin{equation} 
\label{eq:covariancedesipative} 
\mathrm{Cov}(H,S_i^\alpha)
\mathrm{Cov}(H_B,S_i^\alpha)+
\mathrm{Cov}(H_{\mathcal{J}},S_i^\alpha),
\end{equation}
for each spin component $\alpha$, corresponding to the two terms in the spin Hamiltonian in Eq.~\eqref{eq:Bilinearhamiltonian}. 

For the contribution from the Zeeman term $H_B$, we obtain
\begin{equation}
\label{eq:covarianceexpansion}
\mathrm{Cov}(H_B,S_i^\alpha)
=
- \sum_{k\beta} \hbar B_k^\beta
\left(
\langle s_k^\beta S_i^\alpha + S_i^\alpha s_k^\beta \rangle
-2\langle s_k^\beta\rangle \langle S_i^\alpha\rangle
\right),
\end{equation}
For a product state, in particular for the states given in Eq.~\eqref{eq:productcoherent}, Eq.~\eqref{eq:commutatorexpectation} can be used to derive
\begin{equation}
\label{eq:zeemanterm}
\mathrm{Cov}(H_B,S_i^\alpha)
= -\frac{1}{S}\sum_{\beta}
 B_i^\beta G_i^{\beta\alpha},
\end{equation}
where the single-site covariance tensor is defined by
\begin{equation}
\label{eq:convariancetensor}
G_i^{\beta\alpha}
=
\langle \{S_i^\beta,S_i^\alpha\} \rangle
-2 M_i^\beta M_i^\alpha.
\end{equation}

For the bilinear interaction term $H_{\mathcal{J}}$, the corresponding covariance read
\begin{equation}
\label{eq:covarheisebexp}
\mathrm{Cov}(H_{\mathcal{J}},S_i^\alpha)
= 
-\frac{1}{2} \hbar^2 \sum_{kl\mu\nu}
\mathcal{J}_{kl}^{\mu\nu}
\mathrm{Cov}(s_k^\mu s_l^\nu, S_i^\alpha).
\end{equation}
On the manifold of product states, including the product spin-coherent states given by Eq.~\eqref{eq:productcoherent}, all terms in Eq.~\eqref{eq:covarheisebexp} vanish except those for which $k=i$ or $l=i$. By symmetry, we have
\begin{equation}
\label{eq:interactionterm}
\mathrm{Cov}(H_{\mathcal{J}},S_i^\alpha)
=
-\frac{1}{ S^2}\sum_{l \mu \nu}
\mathcal{J}_{il}^{\mu \nu}
G_i^{\alpha\mu} M_l^\nu.
\end{equation}

Combining the two contributions in Eqs.~\eqref{eq:zeemanterm} and~\eqref{eq:interactionterm}, the dissipative covariance term in Eq.~\eqref{eq:magnetizationevolution} is
\begin{equation}
\label{eq:convarianceformofinteration}
\mathrm{Cov}(H,\mathbf{S}_i)
= 
G_i \, \mathbf{B}_i^{\mathrm{eff}}
\end{equation}
with the single-site covariance tensor $G_i$ given in Eq. \eqref{eq:convariancetensor} and the effective magnetic field given in Eq. \eqref{eq:effective_field}.\\


\subsection{Emergence of the classical LLG from QLLG}
By using the product coherent state representation specified in Eqs. \eqref{eq:respsenmagnet} and \eqref{eq:productcoherent}, we get
\begin{equation}
\bra{\psi} \{S_i^\alpha,S_i^\beta\} \ket{\psi}
=
\sum_{\mu \nu}R_i^{\alpha\mu} R_i^{\beta\nu}
\bra{S,S} \{S^\mu_i,S^\nu_i\} \ket{S,S}.
\end{equation}
where
\begin{equation}
U^\dagger(R_i) S_i^\alpha U(R_i)
=
\sum_{\mu}R_i^{\alpha\mu} S^\mu_i,
\end{equation}
such that $R_i \in SO(3)$ is the adjoint rotation matrix. Here, the bipartite correlations become 
\begin{equation}
\begin{aligned}
\bra{S,S} \{S_i^\mu, S_i^\nu\} \ket{S,S}
=&
2\hbar^2 S^2 \delta_{\mu z}\delta_{\nu z}
+
\hbar^2 S(\delta_{\mu\nu} - \delta_{\mu z}\delta_{\nu z}),
\end{aligned}
\end{equation}
where $\delta_{\mu z}$ is the Kronecker delta. By applying the rotation matrices and using the identities 
\begin{equation}
\begin{aligned}
&R_i^{\alpha \beta} \delta_{\beta z} = m_i^\alpha,
\\&
R_i^{\alpha\mu}R_i^{\beta\nu}\delta_{\mu\nu}
=
\delta^{\alpha\beta},\\&
R_i^{\alpha\mu}R_i^{\beta\nu}
\delta_{\mu z}\delta_{\nu z}
=
m_i^\alpha m_i^\beta,
\end{aligned}
\end{equation}
we obtain 
\begin{equation}
\label{eq:keyaverageidentity}
\langle \{S_i^\alpha,S_i^\beta \} \rangle 
=
2\hbar^2 S^2 m_i^\alpha m_i^\beta
+
\hbar^2 S(\delta^{\alpha\beta} - m_i^\alpha m_i^\beta),
\end{equation}
for all components $\alpha,\beta=x,y,z$.
Substituting Eq.~\eqref{eq:keyaverageidentity} into Eq.~\eqref{eq:convariancetensor}, we find
\begin{equation}
G_i^{\alpha\mu}
=
\hbar^2 S(\delta^{\alpha\mu}-m_i^\alpha m_i^\mu).
\end{equation}
This allows the single-site covariance tensor in Eq. \eqref{eq:convarianceformofinteration} to be rewritten as 
\begin{equation}
\label{eq:finalconvaraince}
\mathrm{Cov}(H,\mathbf{S}_i)
= \hbar^2 S
P_i \mathbf{B}_i^{\mathrm{eff}},
\end{equation}
where the projector $P_i$ is defined by
\begin{equation}
P_i^{\alpha\mu}
=
\delta^{\alpha\mu}-m_i^\alpha m_i^\mu,
\end{equation}
and satisfies the identity
\begin{equation}
P_i \mathbf{B}_i^{\mathrm{eff}}
=
-\mathbf{m}_i \times (\mathbf{m}_i \times \mathbf{B}_i^{\mathrm{eff}}).
\end{equation}

Finally, substituting Eq. \eqref{eq:commutator_term} and Eq. \eqref{eq:finalconvaraince} into Eq. \eqref{eq:magnetizationevolution}, we obtain the classical LL equation
\begin{equation}
\dot{\mathbf{M}}_i
=
\tilde{\gamma} \mathbf{M}_i \times \mathbf{B}_i^{\mathrm{eff}}
-  \frac{\lambda}{|\mathbf{M}_i|} 
\mathbf{M}_i \times (\mathbf{M}_i \times \mathbf{B}_i^{\mathrm{eff}}),
\end{equation}
with
\begin{equation}
\tilde{\gamma}=\frac{1}{1+\kappa^2},
\qquad
\lambda=\kappa \tilde{\gamma},
\end{equation}
or equivalently the classical LLG equation given in Eq.\ \eqref{eq:LLGeq}
with $\alpha=\kappa$. Therefore, the QLLG dynamics reduces to the classical 
LLG dynamics on the manifold of pure product spin-coherent states.

\section{Coherent-state manifold dynamics.} \label{sec:largeS_fluctuations} 
To complete the proof that the QLLG dynamics reduces to the classical LLG dynamics in the classical limit, we now demonstrate Step 2 outlined in Section \ref{Sec:dissipativemagnetizationdynamics}. Specifically, we consider QLLG dynamics initialized in a product of spin-coherent states of the form given in Eq.~\eqref{eq:productcoherent} and show that, in the large-spin limit, the state remains asymptotically within the product coherent-state manifold at all times. Namely, we demonstrate that Eq.~\eqref{eq:normmiunit} holds along the QLLG dynamics for the initial state given in Eq.~\eqref{eq:productcoherent}, implying that the magnitude of the local magnetization remains asymptotically constant in the classical limit.

\subsection{Fluctuation Measure}
Using the Casimir identity, 
\begin{equation} 
\sum_{\alpha=x,y,z} (S_i^\alpha)^2 = \hbar^2S(S+1), 
\end{equation} 
one can write  
\begin{equation} |\mathbf M_i(t)|^2 = \hbar^2S(S+1)-V_i(t), 
\label{eq:mean_spin_length} 
\end{equation} 
where 
\begin{equation} V_i(t) = \sum_{\alpha=x,y,z} \operatorname{Var}(S_i^\alpha),\label{eq:Vi} 
\end{equation}
with 
$\operatorname{Var}(O) =  \langle O^2\rangle - \langle O\rangle^2$
for any observable $O$, such as the local spin operator $S_i^{\alpha}$. 

We define the fluctuation measure as
\begin{equation}
V(t) = \max_{1\le i\le N} V_i(t).
\label{eq:Vmax} 
\end{equation}
For the initial product coherent state given in Eq.~\eqref{eq:productcoherent}, Eq.~\eqref{eq:keyaverageidentity} can be used with $\alpha=\beta$ to derive
\begin{equation} 
\begin{aligned}
V(0)&= \max_i V_i(0) \\&= \max_i \frac{1}{2}  \sum_{\alpha= x,y,z}\hbar^2S (1-|m_i^{\alpha}|^2) = \hbar^2 S,
\end{aligned}
\label{eq:V_initial} 
\end{equation}
which indicates that the initial fluctuation measure scales linearly with $S$. To demonstrate Eq.~\eqref{eq:normmiunit}, we show in what follows that the linear scaling in Eq.~\eqref{eq:V_initial} remains valid at all times under the QLLG dynamics.

\subsection{Evolution of the Fluctuation Measure} 
Equation~\eqref{eq:mean_spin_length} implies 
\begin{equation} 
\dot V_i = -\frac{d}{dt}|\mathbf M_i|^2 = -2\mathbf M_i\cdot\dot{\mathbf{M}}_i. \label{eq:Vdot} \end{equation} 
Using Eq.~\eqref{eq:magnetizationevolution}, we obtain 
\begin{equation} \dot V_i = \mathbf M_i\cdot\mathbf A_i + \mathbf M_i\cdot\mathbf F_i, 
\label{eq:Vdot_full} 
\end{equation}
where 
\begin{equation}
\begin{aligned}
\mathbf A_i &=  -\frac{2 i}{\hbar (1+\kappa^2) } \langle[H,\mathbf S_i]\rangle, \\
\mathbf F_i &= \frac{2\kappa}{\hbar(1+\kappa^2)} \operatorname{Cov}(H,\mathbf S_i). 
\end{aligned}
\end{equation}

From the Cauchy--Schwarz inequalities
\begin{equation}
\begin{aligned}
\left|\langle[H,S_i^\alpha]\rangle\right|
&\leq
2\sqrt{\operatorname{Var}(H)\operatorname{Var}(S_i^\alpha)},
\\
\left|\operatorname{Cov}(H,S_i^\alpha)\right|
&\leq
\sqrt{\operatorname{Var}(H)\operatorname{Var}(S_i^\alpha)},
\end{aligned}
\end{equation}
and Eq.\ \eqref{eq:Vi}, one can derive 
\begin{equation}
\begin{aligned}
|\mathbf A_i|=\frac{2}{\hbar(1+\kappa^2)}\left|\langle[H,\mathbf S_i]\rangle\right|
&\leq
\frac{4}{\hbar(1+\kappa^2)}
\sqrt{V_i\operatorname{Var}(H)},
\\
|\mathbf F_i|=\frac{2\kappa}{\hbar(1+\kappa^2)}\left|\operatorname{Cov}(H,\mathbf S_i)\right|
&\leq
\frac{2\kappa}{\hbar(1+\kappa^2)}
\sqrt{V_i\operatorname{Var}(H)}.
\end{aligned}
\end{equation}
Then, Eq.~\eqref{eq:Vdot_full}, together with the triangle and Cauchy--Schwarz inequalities, gives
\begin{equation}
\begin{aligned}
|\dot V_i|
&\leq
|\mathbf M_i|(|\mathbf A_i|+|\mathbf F_i|)\\
&\leq
\frac{2(2+\kappa)}
{\hbar(1+\kappa^2)}
|\mathbf M_i|
\sqrt{V_i\operatorname{Var}(H)}.
\end{aligned}
\end{equation}
Considering $|\mathbf M_i|\le \hbar S$ from Eq.~\eqref{eq:magnetization}, we conclude that 
\begin{equation} \label{eq:upperboundVidot} |\dot V_i| \le \frac{2(2+\kappa)S} {1+\kappa^2} \sqrt{ V_i\operatorname{Var}(H) }. \end{equation}

\subsection{Bound on the Hamiltonian Variance}  

For the Zeeman contribution to the Hamiltonian in Eq.~\eqref{eq:Bilinearhamiltonian}, the $L^2$-triangle inequality gives
\begin{equation}
\begin{aligned} 
\sqrt{\operatorname{Var}(H_B)} &= \left\| H_B-\langle H_B\rangle \right\|_2 \\ &\le \hbar\sum_{i,\alpha}  |B_i^\alpha| \left\| s_i^\alpha-\langle s_i^\alpha\rangle \right\|_2 \\ &= \hbar \sum_{i,\alpha} |B_i^\alpha| \sqrt{\operatorname{Var}(s_i^\alpha)}. \end{aligned}
\end{equation}
Taking into account Eqs.~\eqref{eq:normalizedspin} and ~\eqref{eq:Vmax} we have   \begin{equation} \operatorname{Var}(H_B) \le \frac{C_B}{ S^2} V(t), \label{eq:HBbound} \end{equation} where $C_B=3\left( \sum_i\| \mathbf{B}_i\|\right)^2$. 

Recall that for any bounded operator \(X\) satisfying \(|X|\le \mathcal C\) and any operator \(Y\), such that \(\langle Y\rangle=0\), we have
\begin{equation} \operatorname{Var}(XY) \le \langle X^2 Y^2 \rangle \le \mathcal C^2 \langle Y^2 \rangle = \mathcal C^2 \operatorname{Var}(Y). \label{eq:XYbound} \end{equation} 
Now, consider a single interaction term expressed as \begin{equation} 
s_i^\alpha s_j^\beta = s_i^\alpha \Bigl( s_j^\beta - \langle s_j^\beta\rangle \Bigr) + s_i^\alpha \langle s_j^\beta\rangle,
\label{eq:decomposition}
\end{equation} 
where $\langle s_j^\beta - \langle s_j^\beta\rangle \rangle =0.$ By Eqs. \eqref{eq:normalizedspin} and \eqref{eq:XYbound}, we have 
\begin{align} 
\operatorname{Var} \left( s_i^\alpha (s_j^\beta-\langle s_j^\beta\rangle) \right) &\le \operatorname{Var}(s_j^\beta). 
\label{eq:uppboundvar}
\end{align}
Applying again the \(L^2\)-triangle inequality and using  Eqs.  \eqref{eq:decomposition} and \eqref{eq:uppboundvar} give
\begin{equation}
\label{eq:upperXijbound}
\begin{aligned} \sqrt{ \operatorname{Var}(s_i^\alpha s_j^\beta ) } &\le \sqrt{ \operatorname{Var} \!\left( s_i^\alpha (s_j^\beta-\langle s_j^\beta\rangle) \right) } \\ &\quad + \sqrt{ \operatorname{Var} \!\left( s_i^\alpha \langle s_j^\beta\rangle \right) } \\ &\le 2 \frac{\sqrt{V(t)}}{\hbar S}. \end{aligned} 
\end{equation}
Hence, applying Eq.~~\eqref{eq:upperXijbound} and the \(L^2\)-triangle inequality to the interaction term of the Hamiltonian in Eq.~\eqref{eq:Bilinearhamiltonian} yields 
\begin{equation} \operatorname{Var}(H_{\mathcal{J}}) \le \frac{C_{\mathcal{J}} \hbar^2}{S^2} V(t), \label{eq:HJbound} \end{equation} where \(C_{\mathcal{J}} =9 \left( \sum_{i,j} \|\mathcal{J}_{ij}\|\right)^2 \) . Finally, since \( H=H_B+H_{\mathcal{J}} \), the \(L^2\)-triangle inequality, together with Eqs. \eqref{eq:HBbound} and \eqref{eq:HJbound}, gives the bound on the Hamiltonian variance 
 \begin{equation} \operatorname{Var}(H) \le \frac{C}{S^2} V(t), \label{eq:VarHbound} \end{equation} where $C=(\sqrt{C_B}+ \hbar \sqrt{C_{\mathcal{J}}})^2$ is a constant independent of
$S$.

\begin{figure*}[t]
    \centering
    \includegraphics[width=0.67\textwidth]{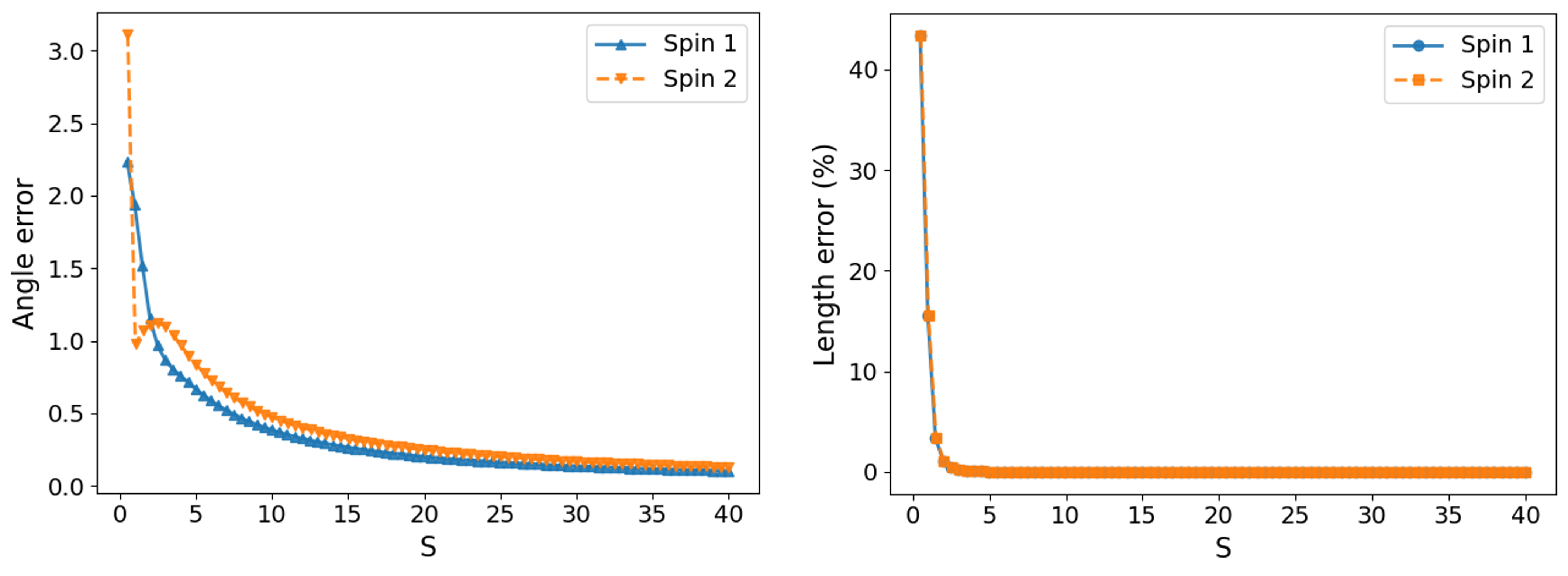}
    
    \caption{The angle and length errors, respectively defined by Eqs. \eqref{eq:angleerror} and \eqref{eq:lengtherror}, quantify the deviations between the magnetizations obtained from the quantum and classical LLG evolutions as a function of the spin number $S$. The angle error is given in radians, while the length error is given as a percentage. }

    \label{fig:Error}
    \end{figure*}

 \begin{figure*}[t]
    \centering
    \includegraphics[width=0.67\textwidth]{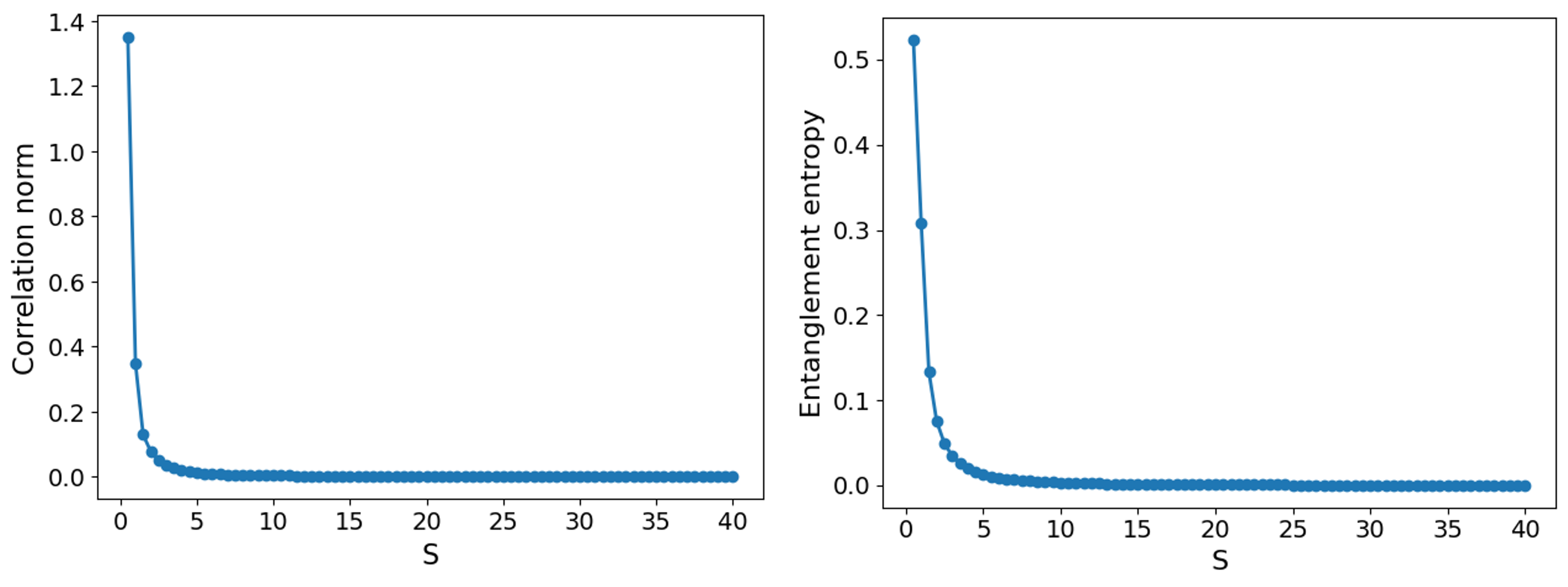}
    
    \caption{Maximum spin-spin correlation norm, $C_{\text{max}}$ given in Eq. \eqref{eq:spinspincorralteion},  and entanglement entropy, $\mathcal{S}_{\text{max}}$ given in Eq. \eqref{eq:engtanglemententropy}, through QLLG dynamics as a function of spin number $S$. The maximum is taken over the intervals $t\in[0,2]$ and $B\in[0,2]$, with time $t$ and magnetic field strength $B$ measured in units of $\hbar/J$ and $J/\mu$, respectively. The plots confirm that quantum correlations vanish in the classical limit at any time and for any magnetic field strength.}
    \label{fig:Correlation}
    \end{figure*}

\subsection{Linear scale of $V(t)$} 
By combining Eqs.  \eqref{eq:upperboundVidot} and \eqref{eq:VarHbound}, we obtain
\begin{equation}
    |\dot{V}_i| \leq  \frac{2(2+\kappa)  \sqrt{C}}{(1+\kappa^2)}  \sqrt{V_i V}.
\end{equation}
Then, taking the maximum of ($i$) and using the definition of the fluctuation measure in Eq.~\eqref{eq:Vmax}, we derive 
\begin{equation}
    |\dot{V}| \leq C' V,
\end{equation}
where $C' = \frac{2(2+\kappa)  \sqrt{C}}{(1+\kappa^2)}$ is a constant independent of $S$.
Gr\"onwall's inequality yields \begin{equation} V(t) \le V(0)e^{C't} = \hbar^2S\,e^{C't}, \label{eq:Gronwall} \end{equation} where the equality follows from Eq.~\eqref{eq:V_initial}. 
 Hence, \begin{equation}
V(t)=O(\hbar^2 S), \label{eq:V_scaling} \end{equation} at any finite time $t$.
 
Eqs.~\eqref{eq:V_scaling} and~\eqref{eq:mean_spin_length} result in 
 \begin{equation} |\mathbf M_i|^2 - \hbar^2S(S+1)= O(\hbar^2 S), 
 \label{eq:finalnormMagnetization}
 \end{equation} 
 or equivalently 
 \begin{equation} \frac{|\mathbf M_i|} {\hbar S} =1+ O( S^{-1}) \label{eq:length_deviation}, \end{equation} which is the same as Eq.~\eqref{eq:normmiunit}. Thus, QLLG dynamics remains asymptotically on the product coherent-state manifold.

It follows from Eq.~\eqref{eq:Gronwall} that Eq.~\eqref{eq:length_deviation} remains valid on finite time scales $t\ll \log S/C'$, and hence the correspondence between QLLG and classical LLG holds in the classical large-spin limit $S$ on the same time scales.

\section{Numerical simulations}

In this section, we numerically verify the analytical quantum-to-classical correspondence established above by directly comparing the QLLG dynamics with the corresponding classical LLG dynamics for a system of interacting spins, from small to large spin quantum numbers $S$.

Let us start by considering a two-spin system described by the following Hamiltonian 
\begin{equation}
H=J
\mathbf s_1\cdot\mathbf s_2-
\mu_{B}
\mathbf B\cdot
(\mathbf s_1+\mathbf s_2),
\end{equation}
in the quantum regime, and the corresponding classical Hamiltonian, where spin operators $\mathbf{s}_i$ are replaced by classical magnetic vectors $\mathbf{m}^c_i$. For simulations, we take Heisenberg exchange $J=1$, Bohr magneton $\mu_{B}=1$, damping rate $\kappa=0.3$, an external magnetic field $\mathbf B= B \left(\frac{1}{3},\frac{2}{3},\frac{2}{3}\right)$ in a fixed direction and variable magnitude $B$, and $\hbar = 1$. The initial quantum state is chosen as a product of two spin-coherent states,
\begin{equation}
|\psi(0)\rangle
=
|\Omega_1\rangle
\otimes
|\Omega_2\rangle,
\end{equation}
parametrized by spherical angles
$\Omega_i \equiv (\theta_i, \phi_i)$, such that quantum and classical magnetization vectors
$\mathbf{m}^q_i = \langle \mathbf{s}_i\rangle$ and $\mathbf{m}^c_i$ are initially identical and point along the unit vector
$\left(
\sin\theta_i \cos\phi_i,\,
\sin\theta_i \sin\phi_i,\,
\cos\theta_i
\right)
$ with the choices
$(\theta_1, \phi_1) = (\pi/3, \pi/5)$ and
$(\theta_2, \phi_2) = (\pi/2, \pi/10)$.

Under the above assumptions, we compare the dynamics of the quantum and classical magnetizations,
$\mathbf{m}_i^{q}$
and
$\mathbf{m}_i^{c}$, by solving the QLLG equation, Eq.~\eqref{eq:shrodingerequation}, and the classical LLG equation, Eq.~\eqref{eq:LLGeq}, respectively. For the classical magnetization we use 
$\mathbf{M}_i^{c}=\hbar S\,\mathbf{m}_i^{c}$, where $\mathbf{m}_i^{c}$ is a unit vector.
Figure ~\ref{fig:Error} illustrates the direction and magnitude errors defined as follows. For direction, we adapt the angle error
\begin{equation}
\label{eq:angleerror}
\mathrm{Angle\ Error}
=
\max_{t,B}
\theta_i,
\end{equation}
where
$\theta_i(t, B)
=
\arccos\left(
\mathbf{m}_i^{q}
\cdot
\mathbf{m}_i^{c}
\right)$
is the instantaneous angle deviation between the quantum and classical spin directions for spin $i$ along their respective dynamics. The length error is evaluated by
\begin{equation}
\label{eq:lengtherror}
\mathrm{Length\ Error}
=
\max_{i, t, B}
(\delta_i),
\end{equation}
such that 
\begin{equation}
\delta_i
=\left | |\mathbf{m}_i^{c}| -|\mathbf{m}_i^{q}| \right |
=\left | 1-
|\langle\mathbf s_i\rangle| \right |.
\end{equation}
The maximums in Eqs. \eqref{eq:angleerror} and \eqref{eq:lengtherror} are taken over the time and magnetic field strength intervals $B\in[0, 2]$ and $t \in [0,2]$. 
Note that the length error converges faster with increasing $S$ than the angular error. The latter quantifies the deviation in the spin directions and is directly related to the phase-space separation of the quantum and classical trajectories. Consequently, the classical approach can be applied even for relatively small values of $S$ while maintaining good accuracy in the spin length. In contrast, achieving a small angular error, and hence reproducing the phase information consistently between the two dynamics, requires substantially larger spin quantum numbers.

\begin{figure*}[t]
    \includegraphics[width=0.67\textwidth]{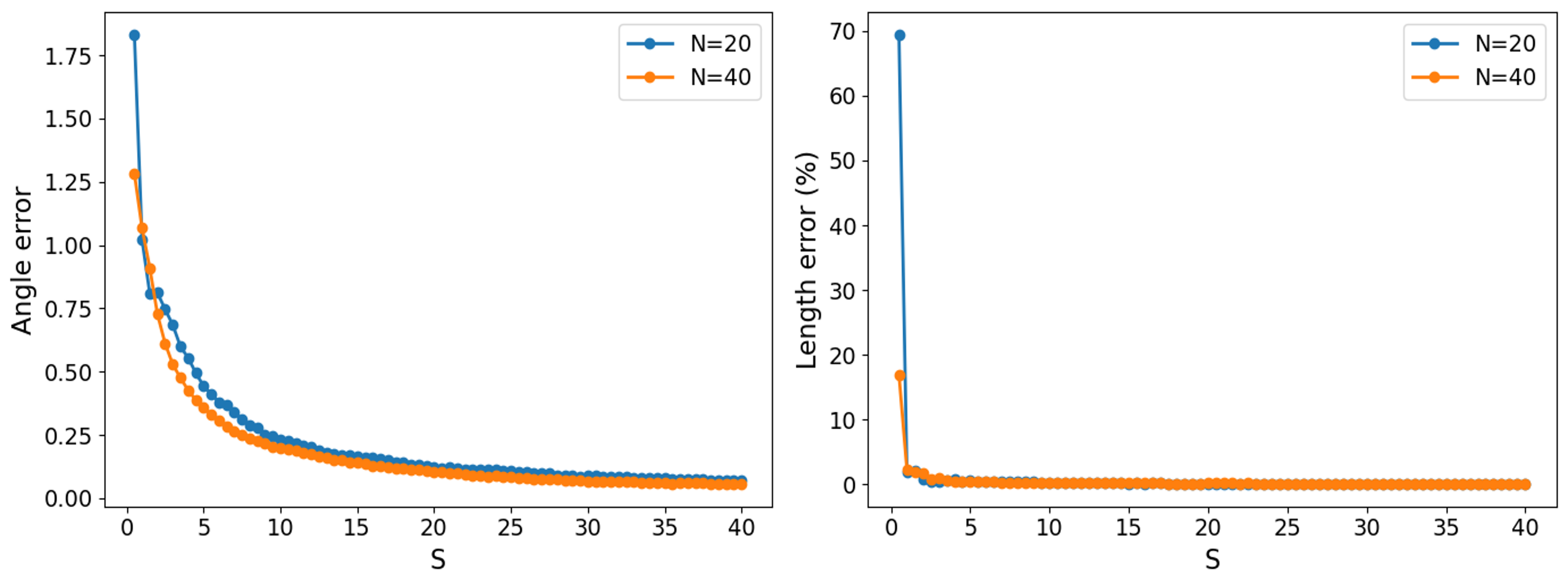}
    \caption{Angle and length errors as functions of the spin quantum number $S$ for Heisenberg chains with $N=20$ and $N=40$ spins, where the angle error given in radians and the length error as a percentage. }
    \label{fig:RBM}
\end{figure*}
 

In Fig.~\ref{fig:Correlation}, we characterize the spin-spin correlations and entanglement. We employ the maximum spin-spin correlation norm
\begin{equation}
\label{eq:spinspincorralteion}
C_{\text{max}} =\max_{t,B} \|C_{\alpha\beta}\|,
\end{equation}
where $C$ is the following correlation tensor
\begin{equation}
C_{\alpha\beta}
=
\langle s_1^\alpha s_2^\beta \rangle
-
\langle s_1^\alpha \rangle \langle s_2^\beta \rangle.
\end{equation}
Entanglement is quantified via the maximum entropy
\begin{equation}
\label{eq:engtanglemententropy}
\mathcal{S}_{\text{max}} = \max_{t,B} S(\rho_1),
\end{equation}
defined through the bipartite von Neumann entropy
\begin{equation}
S(\rho_1)
=
- \mathrm{Tr}\!\left(\rho_1 \ln \rho_1\right),
\end{equation}
where $\rho_1$ is the reduced density matrix of one spin. The maximums are taken over the same time and magnetic field strength interval as stated above. 
In particular, Fig.~\ref{fig:Correlation} shows that quantum correlations are present for small spin quantum numbers, whereas they vanish in the large-spin limit. This highlights that the QLLG dynamics converges to the classical LLG dynamics in the large-$S$ limit, where the product coherent-state structure in Eq.~\eqref{eq:productcoherent} is preserved throughout the QLLG dynamics.

The numerical observations in Figs.~\ref{fig:Error} and \ref{fig:Correlation} verify that QLLG equation asymptotically reproduces the classical LLG dynamics at classical large spin limit. Of the four properties investigated and shown in Figs.~\ref{fig:Error} and \ref{fig:Correlation} the angle stands out as the property that converges most slowly with S, to reache a classical counterpart.

The exact-diagonalization results in Figs.~\ref{fig:Error} and \ref{fig:Correlation} confirm the quantum-to-classical correspondence, established in previous sections, for the interacting two-spin problem. However, our analytical result is valid for any number of spins. To numerically verify this behavior for a larger number of spins, we additionally considered Heisenberg spin chains with \( N=20,40\) spins in Fig. \ref{fig:RBM}.
Here, we use the restricted Boltzmann machine combined with variational Monte Carlo simulations \cite{Carleo2017}  to solve the QLLG equation. The time evolution of the variational parameters was determined using the time-dependent variational principle \cite{CarleoTVDMC}. We consider the average angular error between the quantum and classical magnetizations
\begin{equation}
\label{eq:averageangleerror}
\mathrm{Angle\ Error}
=\frac{1}{N} \sum_i
\theta_i,
\end{equation}
and the average length error 
\begin{equation}
\label{eq:averagelengtherror}
\mathrm{Length\ Error}
= \frac{1}{N} \sum_i
\delta_i,
\end{equation}
where $\theta_i, \delta_i$ are defined as in Eqs. \eqref{eq:angleerror} and \eqref{eq:lengtherror} at fixed time and magnetic field strength $t=2, B=1$.

The results in Figs.~\ref{fig:Error} and Fig.~\ref{fig:RBM} show identical information, with the one exception that Fig.~\ref{fig:Error} is for a two-spin system while Fig.~\ref{fig:RBM}
represents information for chains of 20 and 40 spins. The same qualitative behavior is found for systems with a small and a large number of spins in the Hamiltonian, in the sense that, as function of S, the angle error approaches lower values much slower that the length error.

However, there is one important distinction. In Fig.~\ref{fig:Error}, the angular error is larger for all studied values of $S$ compared with the corresponding values in Fig.~\ref{fig:RBM}. Similarly, the length error drops much more rapidly to small values in Fig.~\ref{fig:RBM} than in Fig.~\ref{fig:Error}.
Taken together, this points to quantum effects becoming less noticeable for systems with a larger number of spins in the Hamiltonian and suggests that quantum technologies explored with quantum spin systems must take this into consideration.

\section{Conclusion}

We have established the quantum-to-classical correspondence for dissipative spin dynamics by deriving the classical LLG equation from QLLG in the classical large-spin limit. This result provides a microscopic foundation for the classical LLG equation and identifies the QLLG equation as a consistent quantum extension of classical magnetization dynamic. This provides a rigorous framework for investigating dissipative quantum spin dynamics in regimes where classical descriptions cease to be valid. The quantum-to-classical correspondence has been demonstrated theoretically and confirmed numerically for spin chains of different size. From the numerical calculations, we also conclude that quantum effects become much less noticeable for systems with a larger number of spins in the Hamiltonian, information that may be relevant for spin-based quantum technologies.

\section{Acknowledgment}
Valuable discussions with Prof. M.I. Katsnelson are acknowledged. O.E. acknowledges financial support from the Swedish Research Council (VR) and the Knut and Alice Wallenberg Foundation (KAW).
O.E. also acknowledges support from the Wallenberg Initiative Materials Science (WISE), funded by the Knut and Alice Wallenberg Foundation, for support, as well as support from STandUPP and eSSENCE. The computations are enabled by resources provided by the National Academic Infrastructure for Supercomputing in Sweden (NAISS), partially funded by the Swedish Research Council (VR).

\end{document}